\documentclass[conference]{IEEEtran}
\IEEEoverridecommandlockouts
\usepackage{cite}
\usepackage{amsmath,amssymb,amsfonts}
\usepackage{graphicx}
\usepackage{textcomp}
\usepackage{xcolor}
\usepackage{multirow}
\usepackage{algcompatible}
\usepackage{multirow}
\usepackage{color, colortbl}

\def\BibTeX{{\rm B\kern-.05em{\sc i\kern-.025em b}\kern-.08em
    T\kern-.1667em\lower.7ex\hbox{E}\kern-.125emX}}
\begin{document}

\title{Interconnect Parasitics and Partitioning in Fully-Analog In-Memory Computing Architectures\\
}


\author{\IEEEauthorblockN{Md Hasibul Amin, Mohammed Elbtity, Ramtin Zand}
\IEEEauthorblockA{Department of Computer Science and Engineering, University of South Carolina, Columbia, SC 29208, USA
}
}

\maketitle

\begin{abstract}
Fully-analog in-memory computing (IMC) architectures that implement both matrix-vector multiplication and non-linear vector operations within the same memory array have shown promising performance benefits over conventional IMC systems due to the removal of energy-hungry signal conversion units. However, maintaining the computation in the analog domain for the entire deep neural network (DNN) comes with potential sensitivity to interconnect parasitics. Thus, in this paper, we investigate the effect of wire parasitic resistance and capacitance on the accuracy of DNN models deployed on fully-analog IMC architectures. Moreover, we propose a partitioning mechanism to alleviate the impact of the parasitic while keeping the computation in the analog domain through dividing large arrays into multiple partitions. The SPICE circuit simulation results for a $400\times120\times84\times10$ DNN model deployed on a fully-analog IMC circuit show that a 94.84\% accuracy could be achieved for MNIST classification application with 16, 8, and 8 horizontal partitions, as well as 8, 8, and 1 vertical partitions for first, second, and third layers of the DNN, respectively, which is comparable to the $\sim$ 97\% accuracy realized by digital implementation on CPU. It is shown that accuracy benefits are achieved at the cost of higher power consumption due to the extra circuitry required for handling partitioning.
\end{abstract}


\section{\textbf{Introduction}}
With the increased computational demands of machine learning (ML) workloads, in-memory computing (IMC) \cite{in-memory-wong} architectures have attracted considerable attention to address the processor-memory bottleneck in conventional von Neumann architectures through executing the logic functions directly on memory via changing the internal memory circuitry. Resistive random access memory (RRAM) \cite{Yin2020HighThroughputIC}, phase-change memory (PCM) \cite{in-memory-PCM}, magnetoresistive random-access memory (MRAM) \cite{ZandVaibhav}, and conductive bridging random access memory (CBRAM) \cite{Molas2019AdvancesIO} are some of the promising technologies that have been utilized in IMC architectures to realize matrix-vector multiplication (MVM) operation in DNNs. While memristive technologies are also leveraged in digital IMC architectures \cite{IMC-Anand-roy, inmemory-fan} to realize logic functions such as XNOR, here we focus on analog IMC architectures due to their great potential for achieving outstanding energy efficiency. For instance, Imec, a world-leading research and innovation hub in nanoelectronics, has recently provided a blueprint towards 10,000 tera operations per second per watt (TOPS/W) DNN inference in \cite{imec2}, which is based on the memristive-based analog IMC architectures. 

Despite the potential benefits of the analog IMC architectures, one of the major factors limiting their wide use in practical ML applications is the large and energy-hungry signal conversion units required to change the computation domain from analog-to-digital (and vice versa) to compute the non-linear vector operations, \textit{e.g.} activation functions in DNNs \cite{PUMA}. Recently, fully-analog IMC architectures are introduced that use memristive technologies to realize both MVM operations and activation functions within the same memory array \cite{IMAC}. These architectures remove the need for the signal conversion unit through maintaining the computation in the analog domain across various layers of DNNs and achieve significant energy and performance improvements. However, due to the fully-analog characteristic of these circuits, interconnect parasitics can have a major impact on the reliability and accuracy of the results obtained by these architectures. Thus, in this paper, we investigate the effect of interconnect parasitics on accuracy of fully-analog IMC architectures and propose an analog partitioning approach to resolve the parasitics effects while keeping the computation in the analog domain.

\section{Fully-Analog IMC Architectures}

\begin{figure}[!t]
\centering
\includegraphics[width=3.4in]{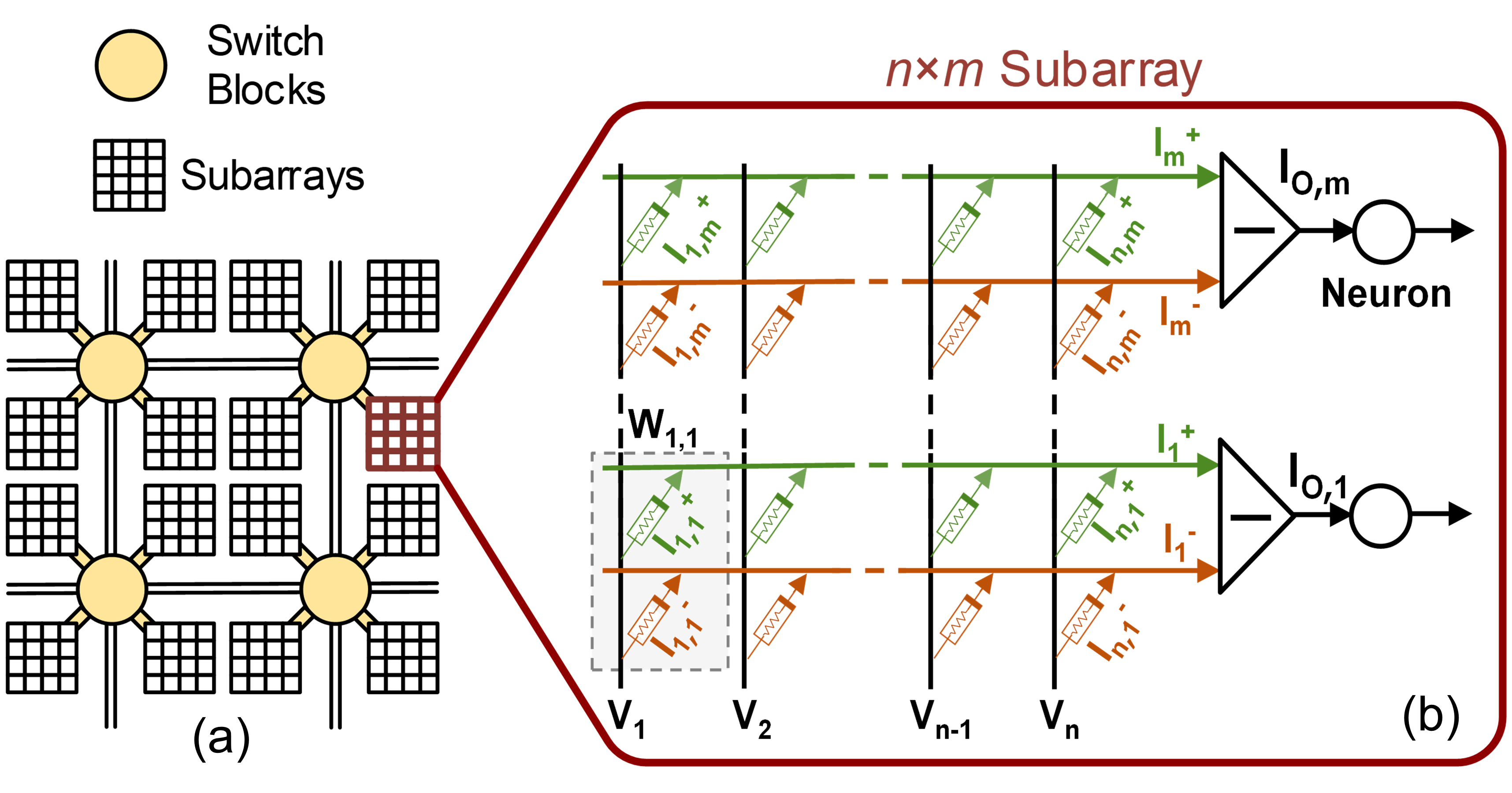}
\vspace{-3mm}
\caption{(a) Fully-analog IMC architecture, (b) An $n \times m$ subarray.}
\label{fig:arch}
\end{figure}

Figure \ref{fig:arch} shows a schematic of the fully-analog IMC architectures, which includes a network of tightly coupled subarrays
interconnected
through programmable switch blocks. Each IMAC subarray consists of memristive crossbar, differential amplifiers, and neuron circuits, as shown in Fig. \ref{fig:arch} (b). The memristor crossbars compute the MVM operation in DNNs in the analog domain through various physical mechanisms such as Ohm's law and Kirchhoff's law in electrical circuits \cite{in-memory-dac}. In particular, the multiplications are performed according to the Ohm's law ($I=GV$), and the accumulation operation is based on the conservation of charge described by the Kirchoff's current law as expressed in the following equation, $I_j = \sum_{j} G_{ij} V_i$, where \textit{G\textsubscript{ij}} is the conductance of the resistive devices between neurons \textit{i} and \textit{j}, $I_j$ is the input current of post-synaptic neuron \textit{j}, and $V_i$ is the output voltage of pre-synaptic neuron \textit{i}.

Each positive and negative weight can be realized through adjusting the relative conductance of two memristive devices that are connected to a differential amplifier. As shown in Fig. \ref{fig:arch} (b), differential amplifiers are connected to two consecutive rows in the crossbar, \textit{i.e.} representing positive and negative weights, and generate an output current of $I_{O,i}=\sum_{k=1}^n (I_{k,i}^+-I_{k,i}^-)$ for the $i$th row, where $n$ is the total number of input nodes in the layer. Whereas $I_{k,i}^+ \propto V_kG_{k,i}^+$ and $I_{k,i}^- \propto V_kG_{k,i}^-$, thus, $I_{O,i} \propto \sum_{k=1}^n V_k(G_{k,i}^+-G_{k,i}^-)$, in which $G_i^+$ and $G_i^-$ are the conductance of resistive devices that are shown in green and red color in Fig. \ref{fig:arch} (b), respectively. Finally, the outputs of the differential amplifiers are connected to the analog neurons to compute the activation functions. In this architecture, each subarray computes both MVM operations and neurons' activation functions in a single DNN layer and passes the result to its downstream neighbor IMAC subarrays that can compute the next layer. 

\section{Interconnect Parasitics Calculation}
\label{sec:parasitic}
\begin{figure*}[t]
\centering
\includegraphics[width=6in]{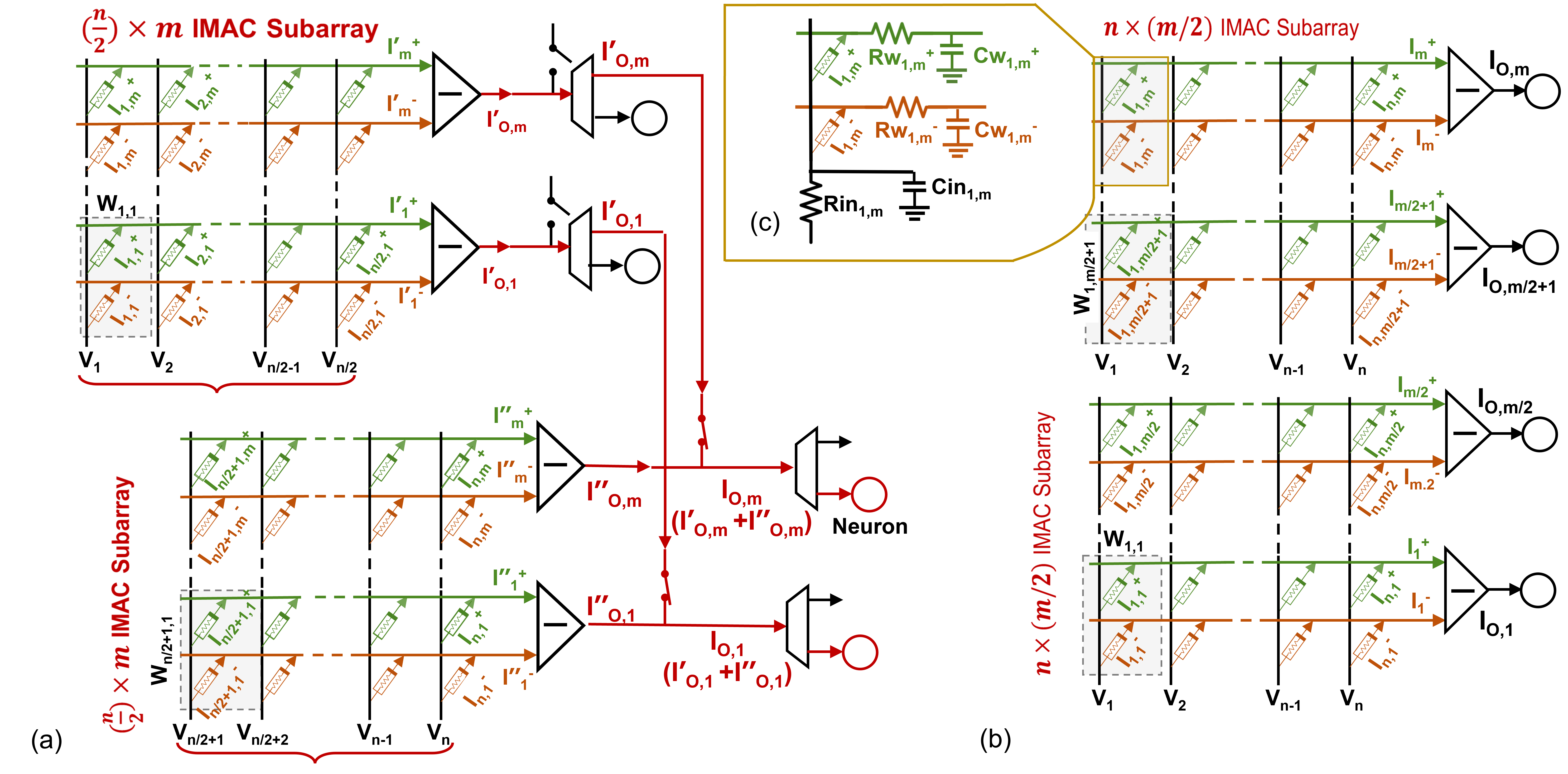}
\vspace{-3mm}
\caption{(a) Horizontal partitioning ($H_P=2$), and (b) vertical partitioning ($V_P=2$) in an analog IMC array. (c) Parasitic capacitance and resistance model.}
\label{fig:partition}
\end{figure*}

Parasitic interconnect resistance ($R_W$) and capacitance ($C_W$) are a function of wire geometry and material properties of interconnections in analog IMC subarrays. Scaling up the size of arrays increases $R_W$ and $C_W$ leading to an increase in the latency of IMC circuits, thus limiting their operating clock frequency. Furthermore, increasing $R_W$ reduces the read margin that can impact the accuracy of analog IMC circuits \cite{Aguirre2020}. Figure \ref{fig:partition}(c) shows the parasitic model for one bitcell in the IMC array. We use the most common equation to find the parasitic resistances for interconnects:

\begin{equation}
R_W = \rho\frac{L}{W.T},
\label{eq:resistance}
\end{equation}

\noindent where $\rho$, $L$, $W$ and $T$ are the resistivity, length, width, and thickness of the metal wire, respectively. Resistivity is commonly a fixed parameter for a specific metal. However, for the sub-micron technology nodes, the resistivity increases due to the surface and grain boundary scattering as the metal width gets comparable to the mean free path of electrons \cite{doi:10.1146/annurev-matsci-082908-145415}. These two well-known scattering effects are quantified using Fuchs-Sondheimer (FS) \cite{fuchs_1938} and Mayadas-Shatzkes (MS) \cite{PhysRevB.1.1382} model respectively.
\begin{equation}
    \frac{\rho_{FS}}{\rho_{Cu}}=1+(1-p)\frac{l_0}{W}
\end{equation}

\begin{equation}
    \frac{\rho_{MS}}{\rho_{Cu}}=\left[1-\frac{3\alpha}{2}+3\alpha^2-3\alpha^3\ln\left(1+\frac{1}{\alpha}\right)\right]^{-1}
\end{equation}

\noindent where $\alpha=\frac{l_0}{d}\frac{R}{1-R}$, $\rho_{Cu}$ is the resistivity of bulk Cu ($1.9\times10^{-9}$ $\Omega m$), $l_0$ is the mean free path of electrons in Cu (39 nm), $W$ is the width of the metal wire, $p$ is the specular scattering fraction, $d$ is the average grain size and $R$ is the probability for electrons to reflect at the grain boundary. $R$ and $p$ are assumed to be 0.3 and 0.25, respectively, and $d$ is assumed to be equal to the wire width as mentioned in the literature \cite{rossnagel},\cite{steinhogl}. The two scattering effects are combined using Matthiessen's rule which results in the following equation \cite{PhysRevB.81.155454}:
\begin{equation}
\begin{aligned}
    \frac{\rho}{\rho_{Cu}} &=1+\left(\frac{\rho_{FS}}{\rho_{Cu}}-1\right)+\left(\frac{\rho_{MS}}{\rho_{Cu}}-1\right)\\
    &=(1-p)\frac{l_0}{W}+\left[1-\frac{3\alpha}{2}+3\alpha^2-3\alpha^3\ln\left(1+\frac{1}{\alpha}\right)\right]^{-1}\\
\end{aligned}
\end{equation}

The parasitic capacitances play major roles in determining the latency of analog IMC circuits. To obtain good accuracy, we fixed the sampling time at 1ns considering the overall latency due to the addition of parasitic capacitances. We use the Sakurai-Tamaru model \cite{1482994} for calculating the parasitic capacitance per length:

\begin{equation}
\label{eq:capacitance}
\begin{aligned}
C_W &=\epsilon\times\frac{1}{2}\left[1.15\left(\frac{W}{H}\right)+2.8\left(\frac{W}{H}\right)^{0.222}\right]\\
            &+\epsilon\times2\left[0.03\left(\frac{W}{H}\right)+0.83\left(\frac{T}{H}\right)-0.07\left(\frac{T}{H}\right)^{0.222}\right]\\
                &\times\left(\frac{S}{H}\right)^{-1.34}
\end{aligned}
\end{equation}

\noindent where $\epsilon=20\epsilon_0$ is the dielectric permittivity of the inter-metal space, $W$ and $T$ are the width and thickness of the metal line, $H=20nm$ is the inter-metal layer spacing and $S$ is the inter-wire spacing \cite{Aguirre2020}. Here, we leverage equations (\ref{eq:resistance}) to (\ref{eq:capacitance}) to model the interconnect parasitics in the SPICE circuit simulations of analog IMC architectures.

\section{Analog Horizontal and Vertical Partitioning}

As the interconnect parasitic impacts can severely degrade the accuracy of the fully-analog IMC circuits, we propose an analog horizontal and vertical partitioning technique to decrease $R_W$ and $C_W$ without requiring to convert the signals from analog domain to digital. Figures \ref{fig:partition} (a) and \ref{fig:partition} (b) provide a schematic of the horizontal and vertical partitioning circuitry, respectively. For the horizontal partitioning, a layer of demultiplexers (DEMUX) is added to the output of the crossbars, which distributes the output currents corresponding to the matrix-vector multiplication results to either neurons in the same subarray for normal non-partitioned operation, or to the next subarray as partial products of that particular partition. Moreover, we locate switches on the output of the crossbars before DEMUX circuits to identify whether the generated output currents should be accumulated with the currents arriving from other subarrays (i.e. partitions) or not. Using these peripheral circuitry and necessary signaling to control the switches and DEMUX circuits, IMC circuit can handle the horizontal partitioning in the analog domain. Figure \ref{fig:partition} (a) shows an example of horizontal partitioning with two partitions $H_P=2$. For vertical partitioning, an $n \times m$ array is divided into multiple $n \times k_i$ subarrays, in which $m=\sum_{i=0}^{V_P} k_i$, where $V_P$ is the total number of vertical partitions. Figure \ref{fig:partition} (b) shows a sample of vertical partitioning with $V_P=2$.

\section{\textbf{Simulation Results and Discussion}}
In this section, we implement a $400\times120\times84\times10$ DNN in SPICE circuit simulator for MNIST \cite{MNIST} handwritten image classification with 20$\times$20 pixels. We use the 14nm High-Performance PTM-MG FinFET model \cite{PTM} along with the $VDD$ and $VSS$ voltages of 0.8V and -0.8V, respectively. Based on the 18nm gate length and the 22nm Fin height of the PTM 14nm FinFET model \cite{ptmpaper}, the layout design parameter $\lambda$ and the metal thickness are fixed to 9nm and 22nm, respectively.

\begin{figure}[!t]
\centering
\includegraphics[width=3.4in]{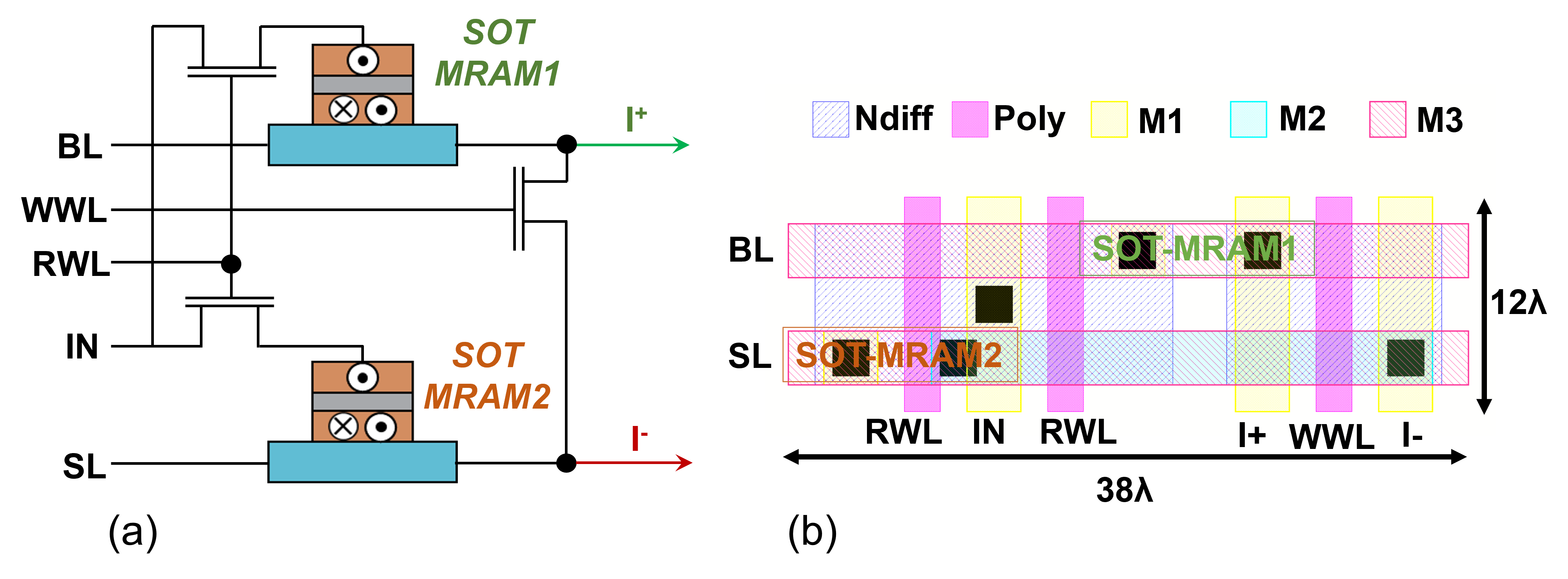}
\vspace{-3mm}
\caption{(a) The SOT-MRAM based synapse bitcell. (b) Layout design.} 
\label{fig:synapse_layout}
\end{figure}

\begin{figure}[!t]
\centering
\includegraphics[width=3.2in]{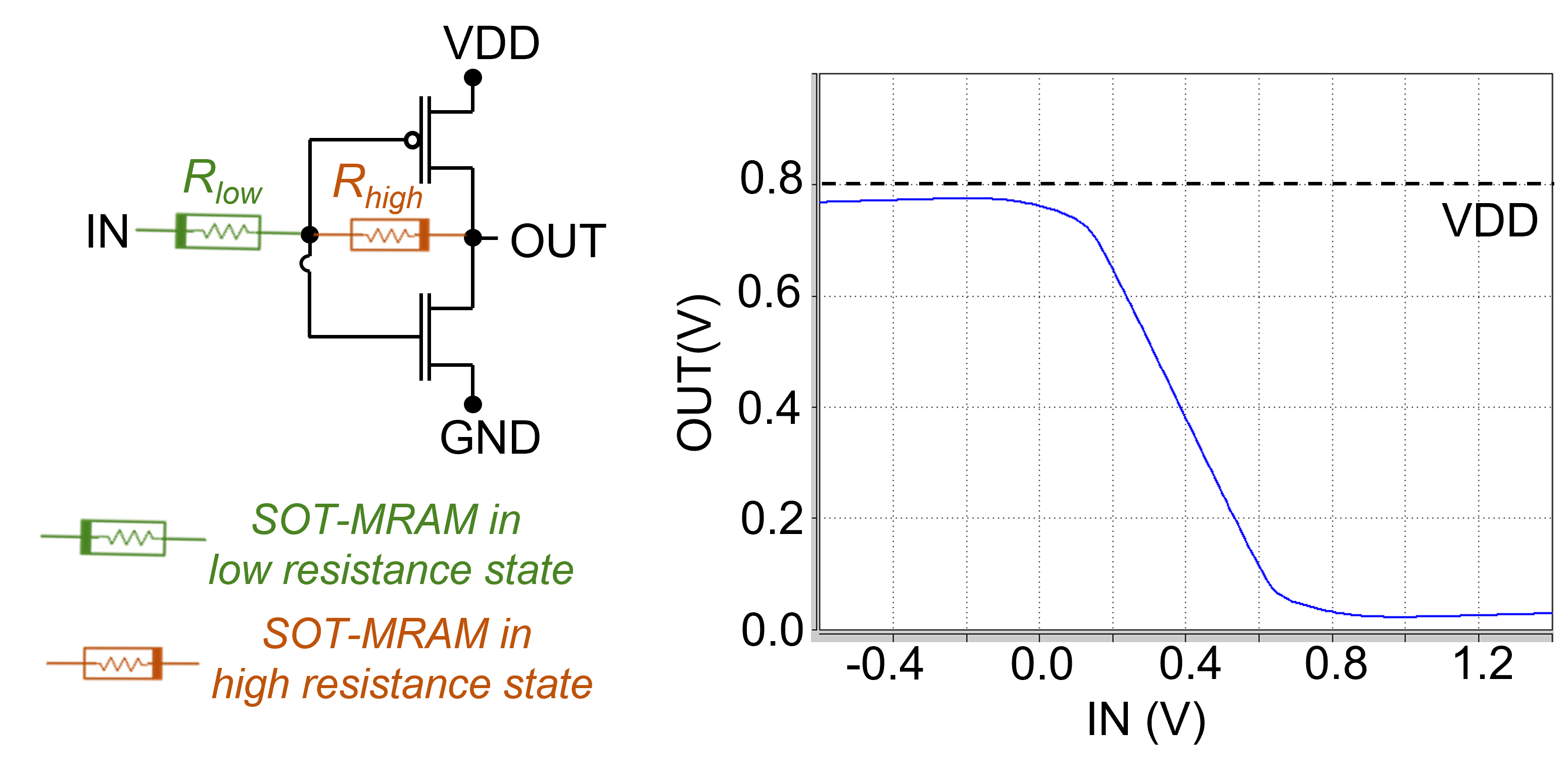}
\vspace{-3mm}
\caption{The memristive sigmoid neuron circuit and SPICE simulation.} 
\label{fig:neuron}
\end{figure}

Here, spin orbit torque MRAM (SOT-MRAM) device model \cite{zand2018fundamentals} is utilized to implement both synaptic structures and activation functions, as shown in Fig. \ref{fig:synapse_layout} and Fig. \ref{fig:neuron}, respectively. We use an analog sigmoidal neuron that includes two resistive devices and a CMOS-based inverter \cite{IMAC}. The resistive devices in the neuron's circuit create a voltage divider that reduces the slope of the linear operating region in the inverter leading to a smooth high-to-low output voltage transition, which enables the realization of a $sigmoid$ activation function.

\begin{table}[htbp]
\caption{Effect of partitioning on the accuracy and power consumption of fully-analog IMC circuits. The dimensions of the layers are L1=400$\times$120, L2=120$\times$84, and L3=84$\times$10.} 
\vspace{-2mm}
\label{tab:partition}
\centering
\begin{tabular}{ccccccccc}
\hline
\multirow{3}{*}{\begin{tabular}[c]{@{}c@{}}Array \\ Size\end{tabular}} & \multicolumn{6}{c}{Number of Partitions}                                                        & \multirow{3}{*}{Accuracy} & \multirow{3}{*}{\begin{tabular}[c]{@{}c@{}}Power\\  (W)\end{tabular}} \\ \cline{2-7}
                                                                       & \multicolumn{3}{c}{Horizontal ($H_P$)} & \multicolumn{3}{c}{Vertical ($V_P$)} &                           &                                                                                       \\ \cline{2-7}
                                                                       & L1          & L2         & L3         & L1         & L2         & L3        &                           &                                                                                       \\ \hline
                                                                      32$\times$32 & 13           & 4            & 3             & 4           & 3            & 1            & 91.71\%                   & 2.640                                                                               \\
                                                                      64$\times$64 & 7            & 2            & 2             & 2           & 2            & 1            & 84.16\%                   & 1.592                                                                              \\
                                                                      128$\times$128 & 4            & 1            & 1             & 1           & 1            & 1            &      15.43\%                      & 0.826                                                                              \\
                                                                      256$\times$256 & 2            & 1            & 1             & 1           & 1            & 1            & 13.17\%                          & 0.829                                                                              \\
                                                                        512$\times$512 & 1            & 1            & 1             & 1           & 1            & 1            & 10.42\%                          & 0.927                                                                              \\ \hline 
\rowcolor[HTML]{EFEFEF}
                                                                      32$\times$32 & 16           & 8            & 8             & 8           & 8            & 1            & \textbf{94.84}\%                   & \textbf{3.375}                                                                                 \\ \hline
\end{tabular}
\end{table}

\begin{figure}[!t]
\centering
\includegraphics[width=3.4in]{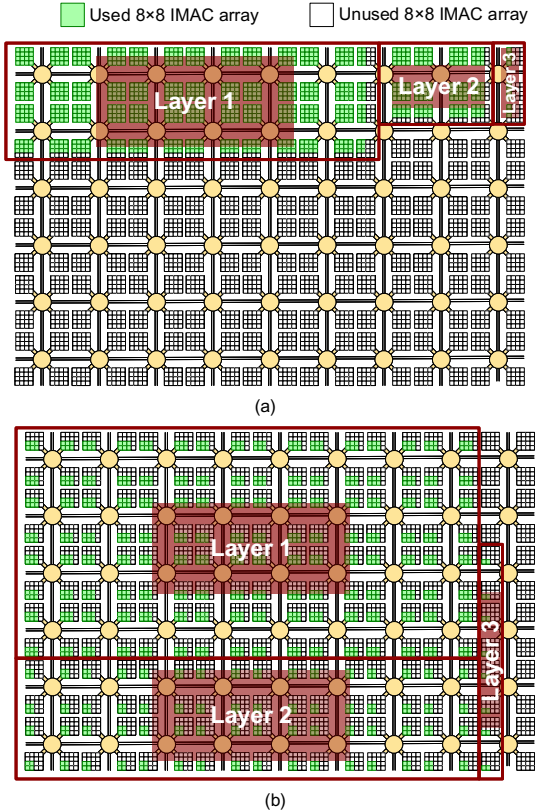}
\vspace{-3mm}
\caption{The deployment of a $400\times120\times84\times10$ DNN on a fully-analog IMC architecture with $32\times32$ subarrays. (a) Maximum subarray utilization with $H_P=[13,4,3]$ and $V_P=[4,3,1]$, (b) Highly-partitioned deployment using $H_P=[16,8,8]$ and $V_P=[8,8,1]$.}
\label{fig:deploy}
\end{figure}

First, we study the effect of partitioning on accuracy and power consumption of the $400\times120\times84\times10$ DNN implemented on a fully-analog IMC circuit. We select the number of partitions for each layer based on various dimensions of IMC subarrays, as listed in Table \ref{tab:partition}. For instance, every layer in the targeted DNN can be deployed on an IMC architecture with $512\times512$ subarrays without any partitioning, while if we use $256\times256$ subarrays, the first layer that includes 400 inputs must be divided into two horizontal partitions to fit into two $256\times256$ subarrays. The results listed in Table \ref{tab:partition} show that without partitioning the deployed model fails to provide reliable classification. It can be seen that as the number of horizontal and vertical partitions increases, the accuracy improves due to the decrease in the length of the interconnects, and consequently their parasitic resistances. However, this is achieved at the cost of increased power consumption due to the extra circuitry added to handle partitioning.

The last row of Table \ref{tab:partition} shows the results for a highly-partitioned case with $H_P=[16,8,8]$ and $V_P=[8,8,1]$ horizontal and vertical partitions for each layer, respectively. This means that assuming an analog IMC architecture with $32\times32$ subarrays, the deployed model does not use the entire capacity of the subarrays, as shown in Fig. \ref{fig:deploy} (b). This deployment scenario results in a high accuracy of 94.84\%, which is close to the $\sim$97\% accuracy realized by the full-precision digital implementations on CPU. However, it is achieved at the cost of higher power consumption and more distributed deployment of DNN model on the architecture that leads to a higher area utilization. 

\begin{figure}[!t]
\centering
\includegraphics[width=3.3in]{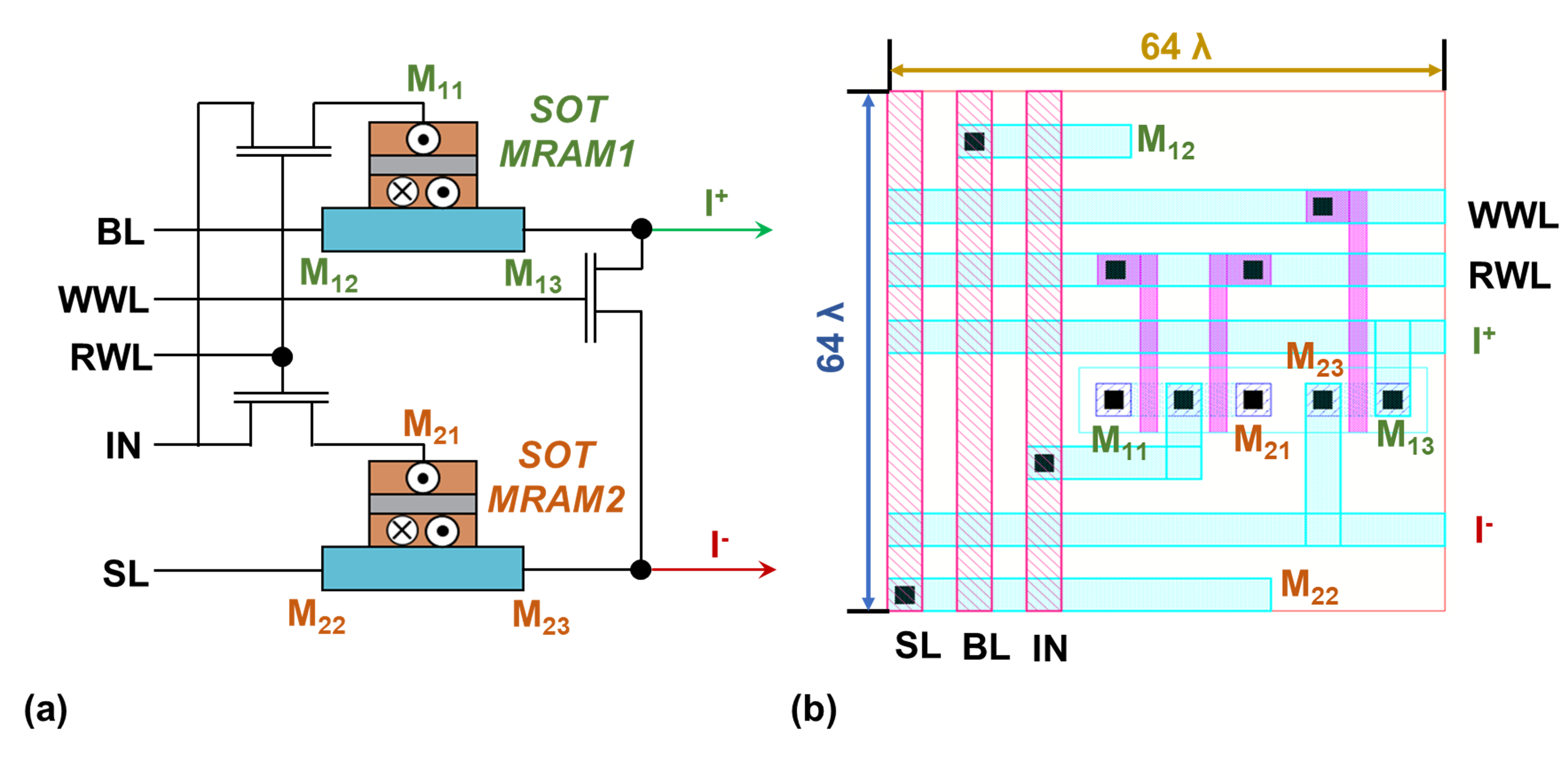}
\vspace{-3mm}
\caption{(a) The SOT-MRAM based synapse bitcell, (b) non-ideal layout design.} 
\label{fig:layout}
\end{figure}

\begin{table}[]
\caption{Effects of partitioning on the deployment of a $400\times120\times84\times10$ DNN on an analog IMC circuit with non-ideal synapse layout design.}
\vspace{-2mm}
\label{tab:layout_nonoptimized}
\centering
\begin{tabular}{ccccccccc}
\hline
\multirow{3}{*}{\begin{tabular}[c]{@{}c@{}}Array \\ Size\end{tabular}} & \multicolumn{6}{c}{Partitioning}                                                        & \multirow{3}{*}{Accuracy} & \multirow{3}{*}{\begin{tabular}[c]{@{}c@{}}Power\\  (W)\end{tabular}} \\ \cline{2-7}
                                                                       & \multicolumn{3}{c}{Horizontal ($H_P$)} & \multicolumn{3}{c}{Vertical ($V_P$)} &                           &                                                                                       \\ \cline{2-7}
                                                                       & L1          & L2         & L3         & L1         & L2         & L3        &                           &                                                                                       \\ \hline
                                                                      32$\times$32 & 13           & 4            & 3             & 4           & 3            & 1            & 73.64\%                   & 1.747                                                                               \\
                                                                      64$\times$64 & 7            & 2            & 2             & 2           & 2            & 1            & 28.44\%                   & 0.926                                                                              \\
                                                                      128$\times$128 & 4            & 1            & 1             & 1           & 1            & 1            &      11.35\%                      & 0.476                                                                              \\
                                                                      256$\times$256 & 2            & 1            & 1             & 1           & 1            & 1            & 11.35\%                          & 0.478                                                                              \\
                                                                        512$\times$512 & 1            & 1            & 1             & 1           & 1            & 1            & 11.35\%                          & 0.479                                                                              \\ \hline 
\rowcolor[HTML]{EFEFEF}
                                                                      32$\times$32 & 16           & 8            & 8             & 8           & 8            & 1            & \textbf{94.04}\%                   & \textbf{2.774}                                                                                 \\ \hline
\end{tabular}
\end{table}

Finally, we investigate the impact of bitcell size on the accuracy and power consumption of the analog IMC circuits. The distance between the wires and the length of the metal lines in an analog IMC subarray depends on the size of the synapse bitcell, which affects the interconnect parasitic resistances and capacitances as described in Section \ref{sec:parasitic}. Figure \ref{fig:layout} shows a non-ideal layout design for the SOT-MRAM based synapse, which leads to a larger bitcell area compared to what is realized in Figure \ref{fig:synapse_layout}. Table \ref{tab:layout_nonoptimized} provides the accuracy and power consumption results for various partitioning scenarios for the $400\times120\times84\times10$ DNN workload deployment on an analog IMC architecture with non-ideal synapse layout design. Accuracy comparisons show that a $\sim$55\% accuracy drop for the non-ideal IMC architecture with $H_P=[7,2,2]$ and $V_P=[2,2,1]$ partitions can reduce to less than 1\% accuracy drop for the highly-partitioned scenario with $H_P=[16,8,8]$ and $V_P=[8,8,1]$ partitions. This shows that increasing the number of partitions can potentially compensate for the imperfections in the layout design at the cost of higher power and area consumption.

\vspace{-1mm}
\section{Conclusion}
Herein, we focused on the impacts of interconnect parasitics on the accuracy of DNN models deployed on the fully-analog IMC architectures. The initial simulation results show that without any mechanisms to resolve the parasitic effects, a $400\times120\times84\times10$ DNN model can barely achieve 15\% accuracy for MNIST classification. Thus, we proposed a horizontal and vertical partitioning mechanism to alleviate the parasitic impacts, while maintaining the computation in the analog domain. This is particularly important in fully-analog IMC architectures which are designed to remove the need for signal conversion units through implementing nonlinear activation functions as well as matrix-vector multiplications in the analog domain. Our proposed partitioning mechanism has shown to be effective to diminish the parasitic impacts such that more than 94\% accuracy could be realized for two different ideal and non-ideal layout design for the IMC circuit.

\bibliographystyle{IEEEtran}
\bibliography{ref}
\end{document}